\documentclass[12pt,a4paper,final]{iopart}

\usepackage{iopams}  
\usepackage{graphicx,cite}
\usepackage[breaklinks=true,colorlinks=true,linkcolor=blue,urlcolor=blue,citecolor=blue]{hyperref}
\usepackage{soul}

\begin{document}

\title{
Strong-field ionization and AC-Stark shifted Rydberg states in OCS}

\author{Mahmoud Abu-samha}
\address{College of Engineering and Technology, American University of the Middle East, Egaila 54200, Kuwait.}
\ead{Mahmoud.Abusamha@aum.edu.kw}

\author{Lars Bojer Madsen}
\address{Department of Physics and Astronomy, Aarhus University, 8000 Aarhus C, Denmark}
\ead{bojer@phys.au.dk}

\begin{abstract}
We present theoretical results for intensity-dependent above-threshold ionization (ATI) spectra from oriented OCS molecules probed by intense femtosecond laser pulses with wavelengths of 800 and 400 nm. The calculations were performed using the time-dependent Schr\"{o}dinger equation within the single-active-electron approximation and including multielectron polarization effects. The results are in qualitative agreement with experimental data [Yu et al., J. Phys. B: At. Mol. Opt. Phys. $\mathbf{50}$, 235602 (2017)]. In particular, characteristic features in the ATI spectra which correspond to resonant multiphoton ionization via highly-excited Rydberg states are captured by the theory.
\end{abstract}

\vspace{2pc}

\section{Introduction}


Intense laser-induced strong-field ionization of the OCS molecule has earned a significant attention in the past decade, where the focus of several experimental works has been on the investigation of the total ionization yields (TIYs) as a function of the orientation angle, $\beta$, of the molecule with respect to the linear polarization of the intense laser pulse ~\cite{HansenJPhysB2011,PhysRevA.89.013405,Johansen_2016,PhysRevA.98.043425}. 
The minimum TIY was obtained at $\beta=0^\circ$, that is when the laser polarization is parallel to the molecular axis. At  $\beta=0^\circ$, suppression of ionization is understood since the laser polarization runs through the nodal plane of the highest occupied orbital (HOMO). The maximum TIY from the HOMO of OCS was reported at $\beta=90^\circ$, that is when the laser polarization is perpendicular to the molecular axis. This measured orientation dependence of the TIYs presented a challenge to ionization models. Neither the molecular Ammosov-Delone-Karinov model~\cite{PhysRevA.66.033402},  the strong-field approximation~\cite{sfa1,sfa2,sfa3}, the weak-field asymptotic theory~\cite{Tolstikhin2011,PhysRevA.87.013406}, the Stark-corrected molecular tunneling theory~\cite{PhysRevA.82.053404,Holmegaard2010}, or  
the Stark-corrected molecular tunneling theory~\cite{PhysRevA.82.053404,Holmegaard2010} could reproduce these experimental results for the OCS molecule.
The behaviour of the measured $\beta$-resolved TIY could, however, be predicted by time-dependent density-functional theory accompanying the experiment in reference \cite{PhysRevA.98.043425}. Recently, we showed that the TIY for OCS as a function of $\beta$ can also be captured by an extension of the single-active-electron (SAE) time-dependent Schr\"odinger equation (TDSE) approach~\cite{PhysRevA.102.063111}. The methodology consists of supplementing the SAE TDSE with an account for the multielectron polarization (MEP) effect as described by the induced-dipole potential, reflecting the laser-induced polarization of the electrons in the cation. With this approach, we have also explained strong-field ionization results in CO~\cite{PhysRevA.101.013433}, as well as in O$_2$, CO$_2$ and  CS$_2$~\cite{PhysRevA.106.013117}; see also the related works by another groups based on very similar approaches~\cite{PhysRevA.95.023407,0953-4075-51-10-105601,PhysRevA.105.023106}. 

So far an investigation of the ability to predict observables that are highly influenced by excited states in strong-field ionization has not been performed in OCS with the effective SAE TDSE approach including MEP effects. Insights into such a possibility is nevertheless highly desirable,  as this could allow one to apply the methodology with confidence for a wider range of observables. Now, experiments published in reference~\cite{Yu_2017} lend us such a possibility for studying excited-state dynamics. In that work, ionization of OCS was considered at different intensities for driving wavelengths of 800 nm and 400 nm. Signature peaks in the intensity-dependent above-threshold ionization (ATI) spectra that did not shift as expected from a direct-ionization model were identified and associated with ionization from excited  states in OCS, in a resonance-enhanced multiphoton ionization (REMPI) process. In this work, we will use the characteristics of the  experimental findings to explore the performance of the SAE TDSE approach including MEP for dynamics that involve excited states. This investigation will, as in the experiment, observe the behaviour of the ATI spectra with varying peak laser intensity. This intensity scan allows us to see effects associated with excited states shifting in and out of resonance in an intensity-dependent manner.
The comparison between our results and the experiment indicates that our SAE potential for OCS including MEP captures the essential physics needed for a description of strong-field ionization involving REMPI mechanisms.


The theoretical and computational models are presented in section \ref{compdet}, followed by results and discussion in section \ref{res} and conclusions in section \ref{conc}. Atomic units are used throughout unless otherwise stated.

\section{Theoretical and Computational Models}
\label{compdet}
In this section, we give a brief overview of the procedures for generating the SAE potential and the wavefunction for the HOMO of OCS, full details are available elsewhere~\cite{PhysRevA.102.063111}. We also provide a brief overview of our TDSE method including the MEP correction; see references \cite{PhysRevA.101.013433,PhysRevA.102.063111,PhysRevA.106.013117} for further description.

The SAE potential for OCS was obtained from quantum chemistry calculations using density-functional theory (DFT), following the procedure described in references \cite{PhysRevA.81.033416,PhysRevA.102.063111}. This potential was expanded in partial waves as $V(\vec{r})=\sum_{l,m=0}^{l_{max}}V_{l0}(r)Y_{l0}(\theta,\phi)$ where $l_{max}=20$. All partial waves in the expansion of $V(\vec{r})$  have $m=0$ since the molecule is linear and the potential is invariant under rotations around the molecular axis. Based on our SAE potential, the HOMO of OCS is a $\pi$ orbital with energy $-0.35$~a.u., which is slightly above a reference orbital energy of $-0.42$ a.u. as obtained from quantum chemistry calculations~\cite{gamess}. The wavefunction corresponding to the HOMO of OCS was obtained using the split-operator spectral method ~\cite{PhysRevA.38.6000}, see reference \cite{PhysRevA.102.063111} for more details. The obtained wavefunction, after being checked for stability on the grid, is used as the initial state for the solution of the TDSE.

The MEP term is approximated as $- \vec{\mu}_{ind}\cdot \vec{r} /r^3 \approx -\alpha_{||} \vec{E}(t)\cdot\vec{r}/r^3$, where $\alpha_{||}$ is the static polarizability of the cation parallel to the laser polarization axis. Reference ~\cite{PhysRevA.106.013117} suggested that there exist a non-vanishing contribution to the induced dipole from the polarizability induced perpendicular to the laser polarization $\alpha_{\perp}$. For the OCS molecule, however, at the orientation angle considered here ($\beta=0^\circ$, corresponding to alignment of the molecular axis with the linear polarization axis), the contribution from $\alpha_{\perp} = (\alpha_{zz}-\alpha_{xx})\sin(\beta)\cos(\beta)$ vanishes. At other orientation angles, our calculations of the induced dipole moment show that  the contribution from $\alpha_{\perp}$ is small and can be neglected. The polarizability $\alpha_{||}$ was computed based on the polarizability components for the OCS$^+$ cation which were reported in our previous work~\cite{PhysRevA.102.063111}. At short range,  a cutoff radius is selected at a radial distance $r_{c}=\alpha_{||}^{1/3}$ such that the MEP cancels the external field at $r\le r_c$~\cite{0953-4075-51-10-105601,PhysRevLett.95.073001,doi:10.1080/09500340601043413}. The reasoning behind this cutoff is that electrons remaining in the cation polarize and thereby set up a field that counteracts the externally applied field. The expression of $r_c$ in terms of $\alpha_{||}$ is determined by setting the sum of the time-dependent interaction potentials equal to zero, $\vec{E}(t)\cdot\vec{r}(1-\alpha_{||}/r_c^3) =0$.

The TDSE is solved for the HOMO of OCS in an effective one-electron potential describing the interaction with the nucleus, the remaining electrons, and the external field, including the induced-dipole MEP term. The TDSE is propagated in the length gauge  with a combined split-operator~\cite{PhysRevA.38.6000}, Crank-Nicolson method; see reference~\cite{Kjeldsen2007a}. The electric dipole field of the laser pulse, $\vec E(t)$, linearly-polarized along the laboratory-frame $z$-axis, is defined in terms of the vector potential $\vec A(t)$ by
\begin{equation}
    \vec{E}(t) = -\partial_t \vec{A}(t) = -\partial_t \left(\frac{E_0}{\omega}\sin^2(\pi t/\tau)\cos(\omega t+\varphi) \right)\hat{z},
    \label{E_field}
\end{equation}
where $E_0$ is the field amplitude, $\omega$ the angular frequency, and $\varphi$ the carrier-envelope phase (CEP) for a laser pulse with duration $\tau$. We consider pulses containing 5 and 8 cycles at angular frequencies of $\omega=0.057$ and 0.114~a.u., corresponding to wavelengths of 800 and 400~nm, respectively. The CEP is kept fixed  ($\varphi=-\pi/2$). Convergence of the results is ensured by using a time-step of $\delta t =0.001$~a.u. in the time propagation and a radial grid containing up to 8192 points and extending to 320 a.u. The size of the angular basis set is limited by setting $l_{max}=60$ in the partial wave expansion of the wavefunction.  The ATI spectra were produced by projecting the wavefunction at the end of the laser pulse on scattering states of the Coulomb potential in the asymptotic region~\cite{Madsen2007}.

\section{Results and discussion}
\label{res}

\begin{figure*}
\includegraphics[width=1.0\textwidth]{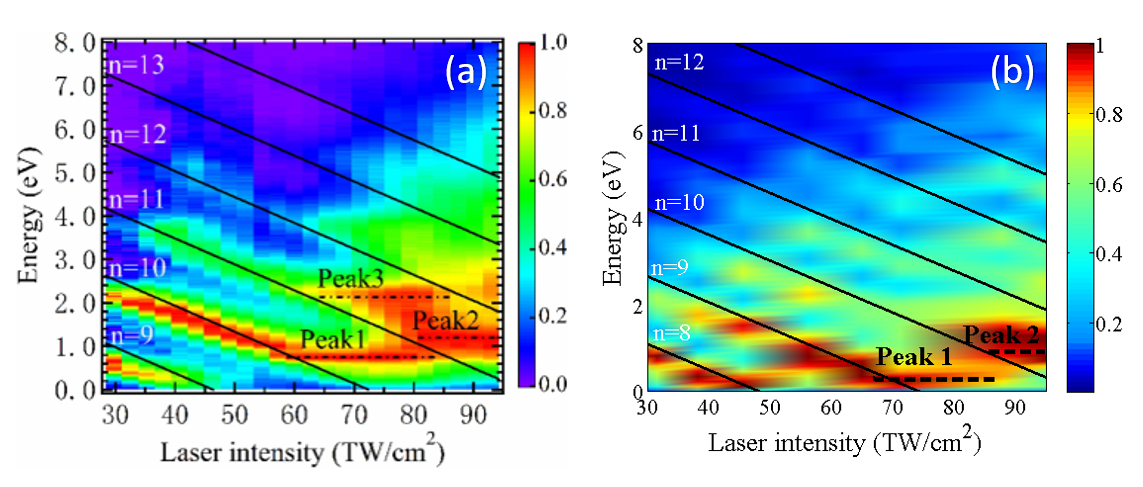}
\caption{\label{fig1} (a) Experimental and (b) Theoretical 2D ATI spectra of OCS at intensities in the range from 28.4 to 94.3 TW/cm$^2$ for 800-nm laser fields. Experimental results are reproduced from reference~\cite{Yu_2017}. The theoretical results were obtained for OCS aligned along the linear polarization direction, i.e., at orientation angle $\beta=0^\circ$. The solid, black lines in (a) and (b) show the kinetic energy positions predicted by equation (\ref{KE}), the integer values $n$ entering that formula are given in (a) and (b). The dashed, black, horizontal lines indicate the position of intensity-independent ATI features (peak1, peak2, and peak3 in (a), and peak1 and peak2 in (b)) associated with resonant ionization.}
\end{figure*}

\subsection{Ionization by 800-nm laser}
In figure~\ref{fig1}~(a) we show intensity-dependent ATI spectra for randomly-oriented OCS from recent experiments~\cite{Yu_2017}. To guide the eye, the spectra are complemented with black, full lines that show how the energy of the ATI peaks would change with laser intensity, assuming the electrons were ionized directly from the HOMO without participation of resonances. The kinetic energy of the photoelectron for each line is calculated using the equation
\begin{equation}
  E_{kin}= n\omega - I_p - U_p,
  \label{KE}
\end{equation}where $n$ is the number of absorbed photons, $I_p$ the ionization potential, and the pondermotive shift $U_p = E_0^2/(4\omega^2)$. The first ATI peak, observed at laser intensities in the range 30$-$45~TW/cm$^2$ corresponds to the OCS molecule absorbing $n=9$ photons (as indicated by the label $n=9$). For the OCS molecule in the intensity range 30$-$60~TW/cm$^2$, the $n=9, 10, 11$, and 12 ATI peaks  shift to lower energy upon increasing the laser intensity, suggesting that at these intensities the HOMO electron is directly ionized into the continuum, without contributions from excited states. At laser intensities exceeding 60~TW/cm$^2$, however, new features appear in the spectra which have been associated with ionization via excited states. These are labeled peak1, peak2, and peak3 in figure~\ref{fig1}~(a) and centered at fixed kinetic energies of 0.7, 1.2 and 2.2 eV. The feature peak1 is clearly visible in the intensity range 60$-$80~TW/cm$^2$. The feature peak3 is a progression of peak1 caused by absorption of an additional photon. These features (peak1 and peak3) have been attributed in reference~\cite{Yu_2017} to (6+4)- and (6+5)-photon resonant ionization through the $^1\Delta$ excited state of OCS: at the considered intensities the HOMO will be excited into the $^1\Delta$ excited state upon absorption of 6 photons and ionized into the continuum upon absorption of additional 4 photons. Regarding the feature labeled peak2, it is fixed at a kinetic energy of 1.2~eV and, compared with peak1 and peak3, appears in the ATI spectra at higher intensities in the range 80$-$95~TW/cm$^2$. This feature (peak2) has been associated with (8+3)-photon ionization through the $^1\Pi$ excited state of OCS~\cite{Yu_2017}. The energy positions of these resonance-mediated features (peak1, peak2, and peak3) are intensity-independent. In reference~\cite{Yu_2017} this characteristic is explained by assuming that the excited resonance states are AC-Stark shifted by the ponderomotive potential, and therefore their energy follow the ponderomotive-potential-shifted ionization threshold, $I_p+U_p$. We note that while the ionization threshold and highly-excited states converging to this threshold are AC-Stark shifted by $U_p$, it is not to be expected that excited states located  as far as $\sim 3, 4$ or 5 photons below threshold follow the simple $U_p$ Stark shift.  Such reasoning could therefore lead to a questioning of the accuracy of the assignment of the relatively low-lying excited states to the REMPI pathways discussed above.

The intensity-dependent ATI spectra of OCS, obtained from our TDSE calculations, are shown in figure~\ref{fig1}~(b). The numerical cost of the extensive scan over laser peak intensities prevents a consideration of laser focal volume and orientation effects. Accordingly, we only show results for OCS with the molecular axis aligned with the linear polarization direction, i.e.,  at  fixed orientation angle $\beta=0^\circ$, and fixed peak intensities. In spite of these restrictions, we find qualitative similarities between our theoretical results (cf. figure~\ref{fig1}~(b)) and the experimental data: at laser intensities in the range 30-60~TW/cm$^2$, the ATI peak positions and their shift with laser intensity suggest that the HOMO electron is directly ionized without participation of high-lying resonances, in agreement with the experimental findings of reference~\cite{Yu_2017}. Moreover,  features associated with resonant ionization via $U_p$ Stark shifted high-lying states, in particular peak1 and peak2, are clearly seen in our theoretical calculations in the same intensity range, 60$-$95~TW/cm$^2$, as reported experimentally.  
The feature peak3 is not well resolved in our TDSE calculations, therefore it will not be discussed here.

Now, a comparison between our theoretical results and the experiment reveal some differences. First of all our SAE model for OCS underestimates the energy of the HOMO: our prediction of the HOMO energy is -0.35~a.u. compared with a literature value of -0.42~a.u., see section~\ref{compdet}. This means that we need less photons to ionize than experimentally, and the observed shift in energy explains the difference in the photon-absorption numbers assigned to the ATI peaks in figure 1 following from the direct ionization model of equation (\ref{KE}). Our results show that it takes a minimum of 8 photons to direct ionization from the HOMO orbital. So, the theoretical direct ATI channels in figure 1(b) are labeled by $n=8, 9, 10, 11, 12$.
Moreover, turning to the resonance features (peak1 and peak2) in figure~\ref{fig1}, the energies of these peaks from TDSE calculations seems to be shifted to lower values in comparison with the experiment: peak1 (peak2) is observed experimentally at electron energy of $\sim 0.7$~eV ($\sim 1.2$ eV), whereas TDSE calculations predict this peak at $\sim 0.5$~eV ($\sim 0.9$ eV).  A possible explanation for this discrepancy is errors in the energies of excited states of OCS predicted by the SAE model. 
This is both  plausible and consistent with our preceding discussion of the SAE performance for OCS, namely that our estimate of the ionization potential for the HOMO orbital being somewhat lower than the literature value. We will come back to the experimental and theoretical excited-state energy positions in section 3.3 below.



\subsection{Ionization by 400-nm laser}

\begin{figure*}
\includegraphics[width=1.0\textwidth]{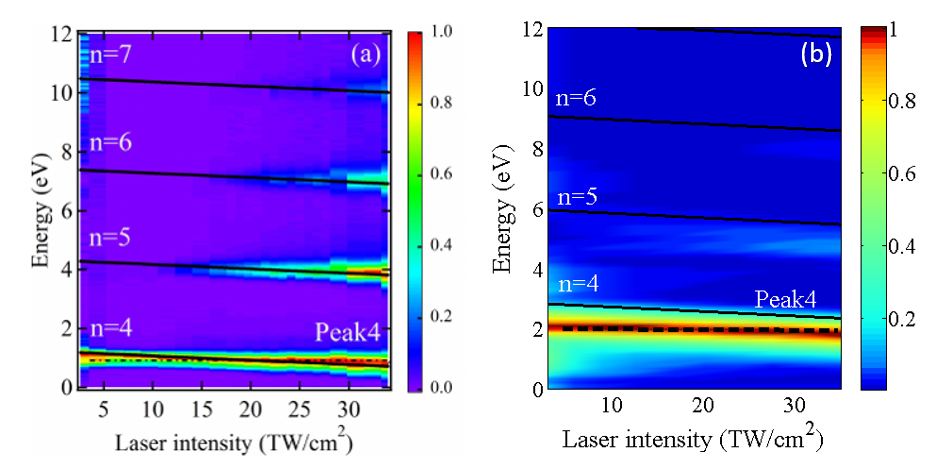}
\caption{\label{fig3} (a) Experimental and (b) Theoretical 2D ATI spectra of the oriented OCS molecule at orientation angle $\beta=0^\circ$ and intensities in the range from 3.5 to 35 TW/cm$^2$ for 400-nm laser fields. The experimental results reproduced from reference~\cite{Yu_2017}. The solid, black lines in (a) and (b) indicate the positions of the ATI peak energies predicted by equation (\ref{KE})
based on direct ionization model without contribution from excited states; the integer values $n$ entering equation~(\ref{KE}) are given in (a) and (b). The dashed lines indicate the positions of intensity-independent ATI features (peak4) associated with resonant ionization via  excited states. }
\end{figure*}



In figure \ref{fig3}(a), we show the experimental 2D ATI spectra of OCS at 400-nm wavelength as obtained from reference \cite{Yu_2017}. From the figure, one can see 4 ATI peaks resulting from absorption of, respectively, $n=4, 5, 6$, and 7 photons. In reference~\cite{Yu_2017}, analysis of contributions from excited states was only considered for the lowest ($n=4$) ATI peak, it was not considered for the higher order ATI peaks ($n=5, 6$, and 7) simply because the ionization yield of these ATI peaks is diminished by going to lower laser intensities. Here, we will follow the same approach and will only focus on the imprint of resonance ionization in the first ATI peak. In figure~\ref{fig3}~(a), the electron kinetic energies calculated based on the direct ionization model of eqaution (\ref{KE}) are shown as solid, black lines. According to the direct ionization model, the $n=4$ ATI peak is expected to shift slightly to lower kinetic energy with increasing intensity. However, at the considered intensity range (3.5$-$34~TW/cm$^2$), the ATI peak seems to be fixed at an intensity-independent energy of about 0.9 eV. This ATI feature 
was labeled peak4 in figure~\ref{fig3}~(a) and its energy position is represented by the dashed, black  line in the figure. This feature was been attributed in reference~\cite{Yu_2017} to (3+1)-photon resonant ionization via the 4p$\sigma$ $\Pi$ Rydberg state.

In figure~\ref{fig3}(b), we show the 2D ATI spectra for OCS as obtained from the present TDSE calculations at a fixed orientation angle of $\beta=0^\circ$. The OCS molecule was probed by 400-nm laser light at intensities in the range 3.5 – 35 TW/cm$^2$. With increasing intensity, our calculations do not show the higher-order ATI peaks ($n=5$, $n=6$, $n=7$) as clearly as  observed experimentally. This is most likely because our calculations were conducted at fixed molecular orientation.
The black solid lines indicate the intensity-dependent positions of the ATI peaks in the case of direct ionization, without contribution from excited states, i.e., the positions predicted from equation (\ref{KE}). Our TDSE calculations of the first ATI peak ($n=4$ peak) show that this peak corresponding to peak4 in figure 2(a) is fixed at an energy of 2.1 eV in particular at laser intensities up to 20~TW/cm$^2$. 
We note that in the experimental measurements, the resonance peak (peak4) was observed at a kinetic energy of 0.9 eV. As discussed in the previous section, a possible explanation for this discrepancy is errors in the energies of excited states of OCS as predicted by the SAE model. This aspect will be discussed further in the following.

\begin{figure*}
\includegraphics[width=0.5\textwidth]{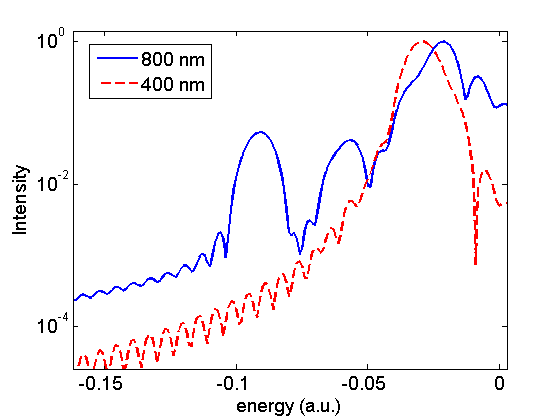}
\caption{\label{fig4} Population in $\Pi$-symmetry excited states of the OCS upon probing the molecule by five-cycle pulse with laser peak intensities of 9~TW/cm$^2$ (at 400 nm; dashed line) and 88~TW/cm$^2$ (at 800 nm wavelength; solid line).}
\end{figure*}

\subsection{Population in excited states and intensity-independent peak features in the ATI spectra }

As can be seen from the experimental results at 800-nm wavelength in figure~\ref{fig1}~(a), the resonance features peak1 and peak2, 
are observed at different intensity ranges. In particular, peak1 
is observed in the intensity range 60$-$80~TW/cm$^2$ and peak2 
in the range 80$-$95~TW/cm$^2$. Regarding the experimental results at 400-nm wavelength in figure~\ref{fig3}~(a), the resonance feature (peak4) is observed at intensities in the range 5$-$15~TW/cm$^2$, see dashed line in figure~\ref{fig3}~(a). 

In the previous section, we established a qualitative agreement between the theoretical intensity-dependent ATI spectra obtained by our TDSE calculations and experimental results. In particular, the ATI features associated with resonance ionization were clearly identified in our calculations.
In the following, we conduct analysis of excited state population based on our TDSE calculations at laser peak intensities where peak2 (at 800~nm) and peak4 (at 400~nm) are observed. To help identify which excited states contribute to ionization at the $\beta=0^\circ$ orientation,  we show in figure~\ref{fig4} population of the excited states of OCS after probing the molecule by five-cycle (800~nm) and eight-cycle (400~nm) pulses with a peak laser intensity of 9~TW/cm$^2$ at 400-nm and 88~TW/cm$^2$ at 800-nm wavelength. At $\beta=0^\circ$, the laser only couples excited states with the same $\pi$ symmetry as the HOMO of OCS. Notice that we use the lower-case letter $\pi$ to describe the symmetry of the active orbital and we will use the wording 'orbital' and not 'state', since the latter is reserved for the many-electron complex. As can be seen from figure \ref{fig4}, when our  OCS  model is probed by the 800-nm pulse, three excited orbitals of $\pi$ symmetry are populated, with the most prominent at an energy of -0.022~a.u. Notice that ionization from this excited orbital by absorption of a single photon results in the first ATI peak at energy of $E_{kin}=-0.022+ \omega~\approx 1.0$~eV. The expected $E_{kin}$ value is in agreement with our theoretical calculations of the peak2 position in figure~\ref{fig1}(b). Notice that the excited orbitals we identify at 800-nm will ionize by a single-photon absorption. However, in reference~\cite{Yu_2017}, the resonance features was assigned to an excited state that would require absorption of additional 3 photons to ionize. This means that this excited-state resonance is located relative far away from the ionization threshold, and it is not obvious that it will satisfy the $U_p$-energy shifting assumed in the data analysis.

In the case when our OCS model is probed by 400-nm pulses with a peak intensity of 9~TW/cm$^2$, the excited orbital at energy of -0.029~a.u. shifts into 3-photon resonance with the HOMO. Ionization by single-photon absorption from this excited orbital results in the first ATI peak at energy of $E_{kin}\approx 2.2$~eV. The expected $E_{kin}$ value from this analysis is in agreement with our theoretical predictions of the $n=4$ position in figure~\ref{fig3}(b). Notice that the results of these analyses are also consistent with the assumption that the AC-Stark shift of the resonantly highly excited orbitals can be well-described by $U_p$. Accordingly, the kinetic energies of electrons ionized from these orbitals remain constant upon increasing the laser intensity as was observed for OCS experimentally~\cite{Yu_2017} and as is now confirmed by the present TDSE results in figures~\ref{fig1} and \ref{fig3}.

\section{Summary and Conclusions}
\label{conc}
Recent experimental work~\cite{Yu_2017} addressed the contribution from excited states to strong-field ionization of randomly oriented OCS at wavelengths of 800 and 400 nm. They associated ATI peaks with ionization from certain excited states. We conducted theoretical analysis of the contribution of excited orbitals to strong-field ionization from the HOMO of the OCS molecule based on TDSE calculations including MEP effects at fixed orientation angle $\beta=0^\circ$. Our theoretical results at both laser wavelengths are in qualitative agreement with the experimental findings. We notice that this agreement is obtained although the present calculations were conducted at fixed molecular orientation of $\beta=0^\circ$ (only $\pi$-symmetry excited orbitals are considered at this orientation) and without accounting for focal-volume averaging. The ability of the present approach to capture essential aspects of excited-state physics under a strong external field, will pave the way for further theoretical investigations of nonlinear processes involving excited molecular states. To mention a few recent applications, this could, for example, include investigations of the role of highly-excited molecular states in analysis of multiphoton-induced free-induction decay in St\"uckelberg spectroscopy ~\cite{PhysRevResearch.4.023135} or in high-order harmonic generation by counterrotating circularly polarized fields~\cite{PhysRevLett.129.173202}.

\section{Acknowledgement}
The time propagation was performed on the cluster of the Centre of Scientific Computing Aarhus (CSCAA). All wavepacket analysis were performed using the Phoenix High Performance Computing facility at the American University of the Middle East (AUM), Kuwait.

\section{Bibliography}
\providecommand{\noopsort}[1]{}\providecommand{\singleletter}[1]{#1}%

\end{document}